%% file: main_asplos.tex
\documentclass[sigplan,screen]{acmart}
\usepackage[normalem]{ulem}

\usepackage{microtype}
\usepackage{xcolor}
\usepackage{url}
\usepackage{subfig}
\usepackage{multirow}
\usepackage{booktabs}
\usepackage{graphicx}
\usepackage{xspace}
\usepackage{algorithm}
\usepackage[noend]{algpseudocode}
\usepackage{balance}
\usepackage{amsmath}
\usepackage{amsthm}
\usepackage{threeparttable}
\usepackage{cleveref}
\usepackage{mathtools}

\newcommand{\argmin}{\operatornamewithlimits{arg\,min}}
\newcommand{\argmax}{\operatornamewithlimits{arg\,max}}

\newcommand{\Sys}{SpotServe\xspace}

\newcommand{\m}{\mathcal}
\newcommand{\revision}[1]{\textcolor{black}{#1}}

\algnewcommand{\LeftComment}[1]{\Statex \(\triangleright\) #1}
\algnewcommand{\LineComment}[1]{\State \textcolor{violet}{\(\triangleright\) #1}}

\makeatletter
\newcommand{\sfgref}[2]{\hyperref[#1]{\ref*{#1}#2}}
\makeatother
\DeclareMathAlphabet{\mathcal}{OMS}{cmsy}{m}{n}

\theoremstyle{plain}

\theoremstyle{definition}

\theoremstyle{remark}

\begin{document}

\title{
\Sys: Serving Generative Large Language Models on Preemptible Instances}

\author{Xupeng Miao}\authornote{Equal contribution.} %
    \affiliation{\institution{Carnegie Mellon University}\city{Pittsburgh}\state{PA}\country{USA}}
    \email{xupeng@cmu.edu}
\author{Chunan Shi}\authornotemark[1]
    \affiliation{\institution{Peking University}\city{Beijing}\country{China}}
    \email{spirited_away@pku.edu.cn}
\author{Jiangfei Duan}
    \affiliation{\institution{The Chinese University of Hong Kong}\city{Hong Kong}\country{China}}
    \email{dj021@ie.cuhk.edu.hk}
\author{Xiaoli Xi}
    \affiliation{\institution{Carnegie Mellon University}\city{Pittsburgh}\state{PA}\country{USA}}
    \email{xiaolix@andrew.cmu.edu}
\author{Dahua Lin}
    \affiliation{\institution{The Chinese University of Hong Kong}\city{Hong Kong}\country{China}}
    \email{dhlin@ie.cuhk.edu.hk}
\author{Bin Cui}
    \affiliation{\institution{Peking University}\city{Beijing}\country{China}}
    \email{bin.cui@pku.edu.cn}
\author{Zhihao Jia}
    \affiliation{\institution{Carnegie Mellon University}\city{Pittsburgh}\state{PA}\country{USA}}
    \email{zhihao@cmu.edu}

\date{}

\input{abstract}

\maketitle
\pagestyle{plain} % should come right after \maketitle

\input{intro}

\input{background}

\input{system}
\input{implementation}

\input{related_work}
\input{conclusion}

\bibliographystyle{plain}
\bibliography{ref}

% \clearpage
% \appendix
% %\input{latency_analysis}
% \input{appendix}

\end{document}

%% file: abstract.tex
%-------------------------------------------------------------------------------
\begin{abstract}
%-------------------------------------------------------------------------------
The high computational and memory requirements of generative large language models (LLMs) make it challenging to serve them cheaply.
%%%
This paper aims to reduce the monetary cost for serving LLMs by leveraging preemptible GPU instances on modern clouds, which offer accesses to spare GPU resources at a much cheaper price than regular instances but may be preempted by the cloud provider at any time.
%%%
Serving LLMs on preemptible instances requires addressing challenges induced by frequent instance preemptions and the necessity of migrating instances to handle these preemptions.

This paper presents \Sys, the first distributed LLM serving system on preemptible instances.
%%%
Several key techniques in \Sys realize fast and reliable serving of generative LLMs on cheap preemptible instances.
%%%
First, \Sys dynamically adapts the LLM parallelization configuration for dynamic instance availability and fluctuating workload, while balancing the trade-off among the overall throughput, inference latency and monetary costs.
%%%
Second, to minimize the cost of migrating instances for dynamic reparallelization, the task of migrating instances is formulated as a bipartite graph matching problem in \Sys, which uses the Kuhn-Munkres algorithm to identify an optimal migration plan that minimizes communication cost.
Finally, to take advantage of the grace period offered by modern cloud platforms, we introduce stateful inference recovery, a new inference mechanism that commits inference progress at a much finer granularity and allows \Sys to cheaply resume inference upon preemption.
% \CA{I think the three aspects need to be changed and can be compressed.}
%%%
We evaluate \Sys on real spot instance preemption traces and various popular LLMs and show that \Sys can reduce the P99 tail latency by 2.4 - 9.1$\times$ compared with the best existing LLM serving systems.
We also show that \Sys can leverage the price advantage of preemptive instances, saving $54\%$ monetary cost compared with only using on-demand instances. The code is open-sourced at: \url{https://github.com/Hsword/SpotServe}.

% \ZJ{Evaluation results.}

\end{abstract}

%% file: intro.tex
\section{Introduction}
Generative large language models (LLMs), such as ChatGPT~\cite{brown2020language} and GPT-4~\cite{openai2023gpt4}, have demonstrated remarkable capabilities of creating natural language texts across various application domains, including summarization, instruction following, and question answering~\cite{zhang2019pretraining, liu2021makes}.
However, the high computational and memory requirements of LLMs make it challenging to efficiently serve them on modern hardware platforms.
To address this challenge, recent work has introduced a variety of approaches to parallelizing LLM inference by partitioning the LLM into multiple sub-models, each of which is deployed on a dedicated GPU.
%%%
For example, GPT-3 includes 175 billion parameters and requires more than 16 NVIDIA A100-40GB GPUs to store the model parameters in single-precision floating points, which costs more than \$66 per hour to serve a single inference pipeline for GPT-3 on AWS~\cite{brown2020language}.
%%%
As the size of LLMs progressively increases, serving them on regular cloud GPU instances becomes prohibitively expensive for most organizations, especially those with limited budgets.

Modern clouds offer a variety of {\em preemptible} GPU instances (e.g., AWS spot instances and Azure spot VMs~\cite{aws_spot, azure_spot}), which provides a more affordable approach to serving LLMs.
%%%
These instances run on spare capacity on modern clouds at a price up to 90\% lower than on-demand instances~\cite{azure_spot}. However, different from on-demand instances, spot instances may be preempted at any time when the capacity is needed by other instances.  
%%%
When a spot instance is preempted, modern clouds provide a {\em grace period} (e.g., 30 seconds for AWS spot instances), which allows the instance to complete running tasks and gracefully stop.

Prior work has introduced several DNN serving systems that leverage spot instances to reduce the monetary cost of DNN inference.
Most of these systems (e.g., MArk~\cite{zhang2019mark}, Cocktail~\cite{276950}) target small DNN models that can fit on a single spot instance with one or multiple GPUs~\cite{kosaian2019parity, yu2022orca}, and handle preemptions using request rerouting~\cite{zhang2019mark} or redundant computation~\cite{kosaian2019parity, bamboo}.
While these approaches can effectively serve small models using data parallelism, they cannot scale to LLMs, serving which requires combining data, tensor model, and pipeline model parallelism~\cite{zheng22-alpa, unger2022unity, piper, miao2023galvatron}.
Model parallelism enlarges the minimal inference granularity from a single GPU instance to a group of instances (i.e., an inference pipeline), which requires more efficient methods to handle preemptions than rerouting and redundant computation, since preemptions are no longer independent and each preemption affects all other instances in the same inference pipeline.

This paper presents \Sys, the first distributed generative LLM serving system on spot instances.
\Sys parallelizes LLM inference across multiple spot GPU instances by combining data, tensor model, and pipeline model parallelism, and produces identical results as serving the LLM using on-demand instances.
Serving LLMs on spot GPU instances requires addressing three main challenges: (1) dynamically reparallelizing LLM inference, (2) cheaply migrating instances, and (3) effectively leveraging grace period.
We elaborate on these challenges and the key ideas \Sys uses to overcome them.

\paragraph{Challenge \#1: dynamic reparallelization.} Serving LLMs requires parallelizing the model parameters and computations across multiple GPUs using a combination of intra-operator (e.g., data and tensor model~\cite{Tensorflow, FlexFlow}) and inter-operator (e.g., pipeline model~\cite{zheng22-alpa, PipeDream}) parallelization strategies. 
%%%
The first challenge \Sys must address is the frequently changing number of available spot instances due to instance preemptions and acquisitions, which requires dynamically adapting the parallelization configuration to achieve optimized LLM serving performance, a problem we called {\em dynamic reparallelization}.
%%%

To address this challenge, \Sys's {\em parallelization controller} dynamically adapts the parallelization strategy for serving LLMs in response to changes in spot-instance availability. \Sys considers both the inference latency of a parallelization strategy and its serving throughput, and uses a hybrid optimization algorithm to balance the trade-off between throughput and latency.
Dynamically reparallelizing LLM inference allows \Sys to quickly adapt to changes to spot instances' availability and requests' arrival rates.

\paragraph{Challenge \#2: instance migration.}
A second challenge \Sys must tackle is minimizing the cost of migrating GPU instances for reparallelization.
%%%
In particular, when transitioning to a different parallelization strategy, \Sys must reinitialize all spot instances to incorporate new model parameters and establish new communication groups.
%%%
Prior work on serving small DNN models on spot instances presumed negligible overheads to reinitialize a spot instance~\cite{276950, zhang2019mark}.
However, we have observed that this assumption is not valid for LLMs, since restarting LLM serving from scratch results in substantial overheads. For example, loading a GPT model with 120 billion parameters from persistent storage takes more than 2 minutes on AWS.

To minimize the migration cost for reparallelization, \Sys opportunistically reuses the model parameters and intermediate results such as key/value cache of an inference request (see \Cref{sec:background}) to avoid unnecessary communication between instances.
The task of mapping available spot instances to the device mesh of a parallelization strategy is formalized as a {\em bipartite graph matching} problem in \Sys, which leverages the Kuhn-Munkres (KM) algorithm to identify an optimal device mapping that minimizes the cost of migrating spot instances for reparallelization.
%%%
In addition, to decide in which order to migrate instances, \Sys's {\em migration planner} leverages the sequential execution order of pipeline stages to overlap instance migration with inference computation.

\paragraph{Challenge \#3: grace period.}
%%%
Leveraging the grace period provided by modern clouds presents another challenge as the inference time for LLMs may surpass the grace period, therefore leading to unfinished requests.
%%%
In existing spot-instance serving systems, these unfinished requests are generally rerouted to other inference pipelines, where the inference computation of these requests is restarted from the beginning.
%%%
This approach does not efficiently use grace period and results in redundant computations.

To take advantage of grace period, \Sys leverages the {\em autoregressive} nature of LLMs and introduces {\em stateful} inference recovery, which allows inference engines in \Sys to commit their progress at the token level, rather than the request level as seen in prior work. \Sys's inference engine uses a \textit{just-in-time} (JIT) arrangement to determine when to migrate the key/value cache of committed tokens to other available instances, which use the cached results to resume inference.
% \ZJ{@Xupeng: elaborate our stateful inference mechanism}

%\paragraph{Evaluation.}
The above techniques allow \Sys to significantly outperform existing approaches.
We have evaluated \Sys on real traces and a variety of LLMs and shown that \Sys reduces the P99 tail latency by 2.4 - 9.1$\times$ compared with existing LLM serving systems. 
In addition, \Sys can utilize spot instance to reduce the monetary cost for serving LLMs by up to $54\%$ compared with the on-demand instance while preserving close average inference latency.

%% file: background.tex
\begin{figure*}[t]
    \subfloat[Incremental decoding in generative LLM]{
        \centering
        \includegraphics[width=0.34\linewidth]{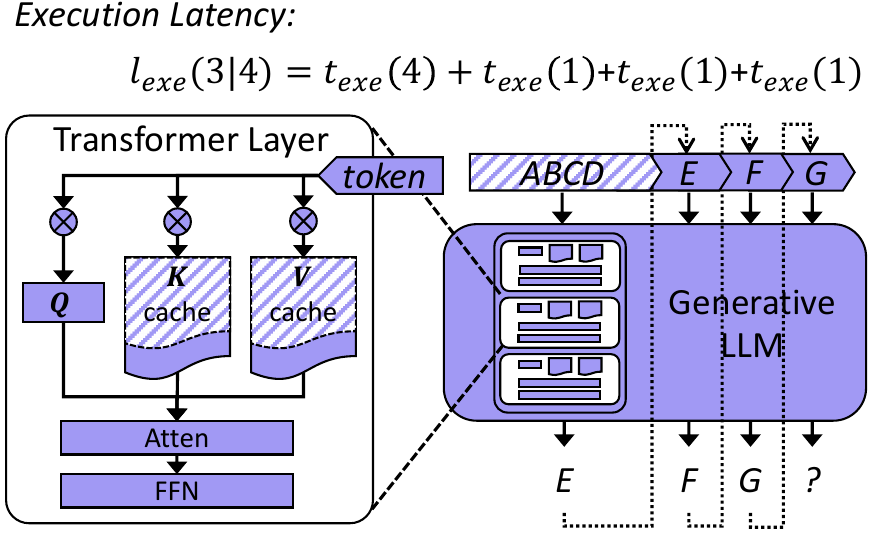}
        \label{fig:decoding}
    }%
    \subfloat[Distributed LLM inference (P=2, M=2, batch size=1) on preemptible instances]{
        \centering
        \includegraphics[width=0.63\linewidth]{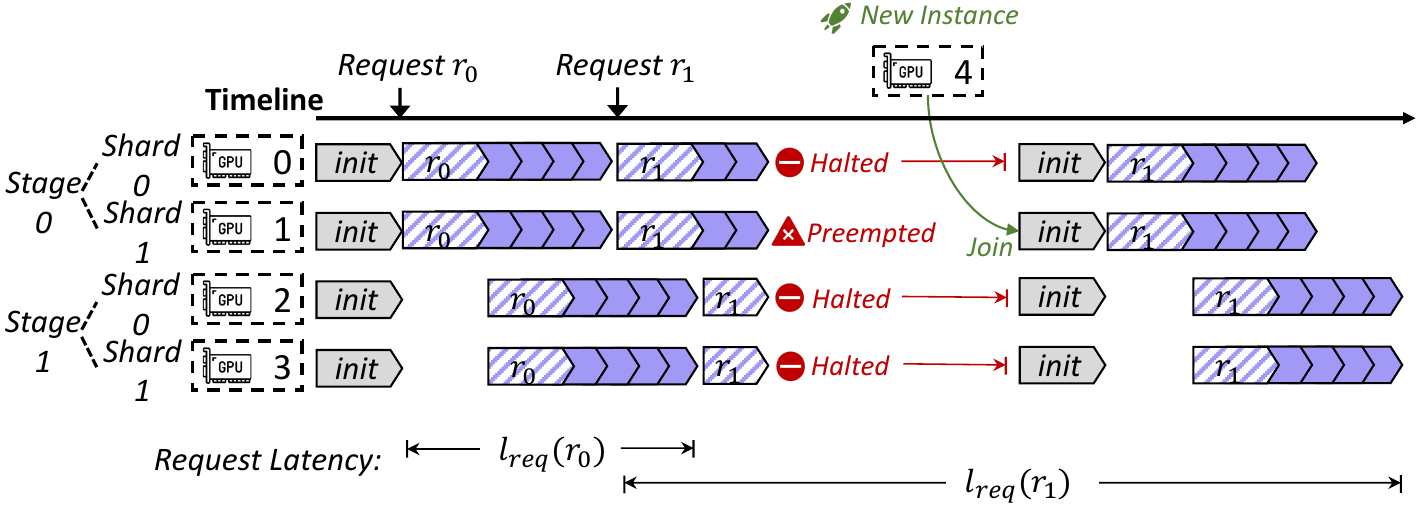}
        \label{fig:preemp}
    }
  \caption{Illustration of incremental decoding in generative LLM and distributed LLM inference on preemptible instances}
\end{figure*}

\section{Background and Related Work}
\label{sec:background}

\subsection{Generative LLM Inference}
Generative LLMs usually stack several identical Transformer~\cite{Transformer} layers and each layer is mainly made up of multi-head attention mechanisms and feed-forward networks (FFNs), as shown in Figure~\ref{fig:decoding}.
The generative LLM adopts the auto-regressive decoding mechanism, leading to an incremental inference process consisting of several iterations.
We dive deeper into the iterative process to provide a better understanding of the generative LLM inference. 
Given a batch of input requests, the corresponding execution latency $l_{exe}$ is divided into two components in E.q.\eqref{eq:exe}:
\begin{align}
    l_{exe}(S_{out}|S_{in}) & = t_{exe}(S_{in}) + \sum_{i=1}^{S_{out}}t_{exe}(S_{in}+i)\label{eq:exe}\\
    & \approx t_{exe}(S_{in}) + S_{out}\times t_{exe}(1)\label{eq:cache}
\end{align}

\noindent where $t_{exe}$ indicates the LLM's execution time cost as a function of decoding sequence length, $S_{in}$ is the sequence length of the input tokens provided the users, and $S_{out}$ is the sequence length of output tokens the generated by the LLM. The first iteration is the \textit{initial phase}, which takes all input tokens, processing them in parallel, and produces the first output token. After that, each \textit{incremental decoding} iteration considers all input together with currently generated tokens and generates one output token.
Figure~\ref{fig:decoding} illustrates an example where the generative LLM takes ``\textit{ABCD}'' as the input sequence (i.e., $S_{in}=4$) and generates one output token in each iteration.

Existing generative inference systems (e.g., FasterTransformer~\cite{fastertransformer}, Orca~\cite{yu2022orca}, FairSeq~\cite{DBLP:conf/naacl/OttEBFGNGA19}, Megatron-LM~\cite{Megatron}) use a key-value (KV) caching optimization that caches the keys and values of all Transformer layers in GPU device memory.
The KV cache helps avoid recomputing preceding tokens during attention calculation, resulting in an almost constant per-iteration overhead (i.e., $t_{exe}(1)$ in E.q.\eqref{eq:cache} and Figure~\ref{fig:decoding}).
However, as the output sequence grows longer, the memory space of KV cache keeps expanding, which can be huge in real workloads (i.e., 1.7 GB per-sequence in LLaMA-13B \cite{vLLM}, or even terabytes in OPT-175B~\cite{DBLP:journals/corr/abs-2303-06865}).

\begin{figure}[t]
    \subfloat[Pipeline model parallelism]{
        \centering
        \includegraphics[width=0.54\linewidth]{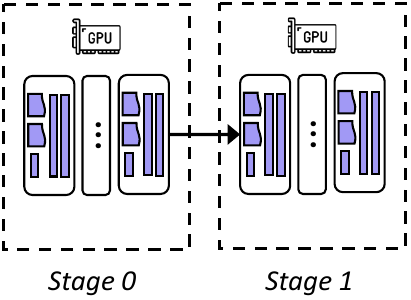}
        \label{fig:parallel_pp}
    }%
    \subfloat[Tensor model parallelism]{
        \centering
        \includegraphics[width=0.45\linewidth]{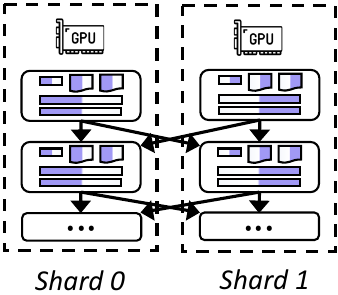}
        \label{fig:parallel_tp}
    }
  \caption{Illustration of different model parallelisms}
  \label{fig:parallel}
\end{figure}

\subsection{Distributed Inference of DNNs}
\label{sec:dist_infer}
Existing distributed DNN serving systems such as NVIDIA Triton~\cite{nvidia_triton} generally maintain multiple concurrent inference pipelines, each of which independently serves an inference engine such as FasterTransformer~\cite{fastertransformer} on several GPU devices. 
An inference server receives input requests, partitions them into small batches, and dispatches them to these inference pipelines.
All GPUs of each inference pipeline work collaboratively to perform DNN inference and send the output back to the inference server.
For each inference request, its end-to-end inference latency $l_{req}$ is divided into two parts: the scheduling overhead $l_{sch}$ and the execution latency $l_{exe}$. The former is determined by the arrival rate of input requests and the peak serving rate of the inference system. 
If the arrival rate exceeds the peak serving rate, input requests cannot be processed in time, resulting in an increase of scheduling overhead. 
In this case, the inference system must improve the serving capability by improving the overall throughput.
When the arrival rate is lower than the peak serving rate, $l_{sch}$ still exists because the requests' arrival intervals can be non-uniform, in which case burst requests introduce scheduling overheads.
All GPUs within an inference pipeline parallelize inference computation by combining two categories of parallel paradigms, as illustrated in Figure~\ref{fig:parallel}.

\paragraph{Inter-operator parallelism.} Pipeline model parallelism~\cite{gpipe} is the most representative inter-operator parallelism strategy, which groups operators into stages with data dependencies. Figure~\ref{fig:parallel_pp} shows an example of partitioning the model into two stages and each stage has half consecutive Tranformer layers. These stages can form a pipeline based on certain pipeline scheduling mechanisms~\cite{PipeDream,pipedream-2bw} that brings stage overlapping as well as cross-stage communications.

\paragraph{Intra-operator Parallelism.} Tensor model parallelism~\cite{Megatron} splits each DNN operator into several shards across the devices.
As shown in Figure~\ref{fig:parallel_tp}, the corresponding tensors are also sharded based on certain distributed data layout. 
The participating devices compute in parallel and perform collective communications (i.e., All-Reduce) to transform the data layout if necessary. 

Note that both the data dependencies in pipeline model parallelism and the collective communications in tensor model parallelism do not naturally provide fault tolerance.
The preemption of a single GPU instance can potentially hang all the other instances in the same inference pipeline.
A preemption may also potentially break multiple inference pipelines if these pipelines are supported by different GPUs located on the same preempted instance.
This chain crashing problem enlarges the affects of a single instance's preemption from itself to several pipelines. 
The affected instances are not physically terminated but stay idle until new instances are allocated to establish new inference pipelines.

\subsection{Preemptible LLM Inference}
Recent work has introduced a variety of techniques on how to handle instance preemptions when using cheap spot instances for DNN computation.
For example, Varuna~\cite{athlur2022varuna} maximizes training throughput by dynamically changing the hybrid data and pipeline parallel configuration after each instance preemption.
Bamboo~\cite{bamboo} uses a redundancy-based preemption recovery mechanism in pipeline parallel training by replicating each instance's computation on another spot instance. 
However, these techniques are designed for distributed DNN training and do not apply to generative LLM serving.
Since distributed LLM inference is an important and timely research topic, accompanied by huge emerging demands in practice, it is obvious that serving LLM over spot instances could be a worthwhile attempt.
Existing LLM serving systems such as FasterTransformer~\cite{fastertransformer} do not provide any preemption handling capability for distributed LLM inference.

Figure~\ref{fig:preemp} illustrates the obstacle of existing systems when serving LLMs on preemptible instances.
We show an example of one distributed LLM inference pipeline deployed over 4 instances (one GPU per instance) through the combination of 2-way pipeline model parallelism and 2-way tensor model parallelism.
The inference process starts after system initialization and it goes well for request $r_0$. But during the incremental decoding process of request $r_1$, GPU 1 unfortunately gets preempted at a certain timestamp and the other three GPUs has to be halted in the meanwhile. Due to the preemption, the inference state of $r_1$ (i.e., KV cache) is lost. When a new GPU instance is launched, it can join the inference pipeline and these 4 GPUs reinitialize and restart the inference process of $r_1$.
As a result, the request latency can be significantly increased due to instance preemption handling.

%% file: system.tex
\section{\Sys Design}
The increased request inference latency caused by instance preemption is mainly manifested in three aspects. 
Firstly, once a preemption happens, the entire inference pipeline comes to a halt, which may result in request waiting overhead and/or additional request scheduling overhead (i.e., rerouting to another inference pipeline).
Secondly, after a new instance joins, there are necessary system initialization costs, such as launching the distributed inference engine and loading model parameters.
Finally, throughout this process, the overall reduction in system throughput can potentially lead to an accumulation of subsequent incoming requests, thereby amplifying their inference latency.

We develop \Sys to mitigate the impacts of these issues on the end-to-end inference latency. First, to alleviate the waiting time caused by the integration of new instances, \Sys facilitates the integration of on-demand instances to ensure swift instance acquisition.
Second, to reduce the runtime overhead of system re-initialization, \Sys introduces an efficient context management mechanism that leverages inter-instance network links to preserve inference progress (in the form of KV cache) and obviate the need for expensive model parameter reloading.
Third, to strike a better balance among serving throughput, latency, and monetary cost during node availability fluctuations, \Sys incorporates a workload-aware adaptive configuration optimization algorithm, which dynamically selects an optimal parallel configuration, enabling real-time dynamic context migration and seamless configuration transitions.

\subsection{System Overview}
\begin{figure}
    \centering
    \includegraphics[width=1.0\linewidth]{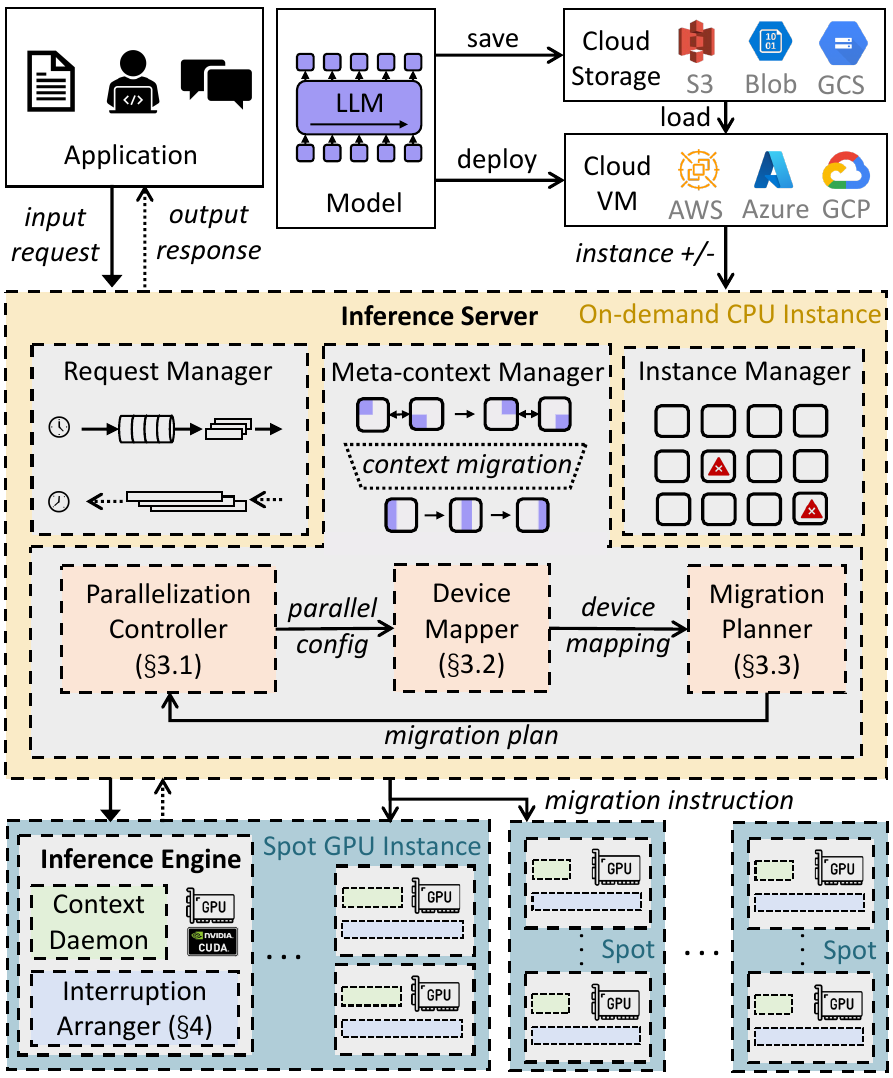}
    \caption{An overview of \Sys.}
    \label{fig:overview}
\end{figure}

Figure~\ref{fig:overview} illustrates an overview of \Sys. The inference server is deployed on a dedicated on-demand CPU instance and hosts a request manager, a meta-context manager, and an instance manager.
The \textit{request manager} is responsible for receiving input requests, dynamically partitioning them into batches, assigning these batches to inference instances running on spot GPU instances, and ultimately collecting generated outputs from the inference instances, sending the results back to users.
%%%
The \textit{instance manager} interacts with the cloud and receives instance preemption/acquisition notifications.

\Sys's inference engine is deployed on each spot or on-demand GPU instance to serve LLM inference.
Each inference engine includes a \textit{context daemon} that manages the model parameters (i.e., model context) and intermediate activations (i.e., cache context) for different requests inside a certain GPU.
The inference engine can access these context information through the proxy provided by the context daemon.
If the inference engine has to be interrupted due to the preemption of dependent instance, the context daemon process is still alive and avoids to reload the context into GPU when restarting inference.

When the system's serving capability becomes incompatible with the workload or is about to, the \textit{meta-context manager} manages the adjustment of the parallel configuration by sending instructions for context migration to all GPU instances. The new configurations are proposed by the \textit{parallelization controller} and materialized by the \textit{device mapper} and \textit{migration planner}. Each inference engine also launches an \textit{interruption arranger} to support stateful inference recovery for lower inference latency.

For the rest of this paper, we first introduce the \Sys's design, including parallelization controller in \S\ref{sec:cfg_opt}, device mapper in \S\ref{sec:mapper}, migration planner in \S\ref{sec:migrate}, and interruption arranger in \S\ref{sec:inter_infer}. Finally, we introduce \Sys's  implementation in \S\ref{sec:impl} and evaluate its performance in \S\ref{sec:eval}.

\subsection{Parallelization Controller}
\label{sec:cfg_opt}

\begin{algorithm}[t]
\small
\begin{algorithmic}[1]

\Function{ConfigOptimizer}{$N_t, C_t, \alpha_t$}
    \If{$\exists C. \phi(C) \geq \alpha_t$ \textbf{and} \text{cloud has enough instances for} $C$}
    \State $C_{t+1}\gets \argmin_{C|\phi(C) \geq \alpha_t}l_{req}(C)$
    \Else
    \State $C_{t+1}\gets \argmax_{C|N_t} \phi(C)$
    \EndIf
    \State $\Delta \gets \texttt{\#Instances}(C_{t+1})-N_t$
    \If{$\Delta>0$}
    \State $\texttt{InstanceManager.alloc}(\Delta, \textit{ondemand\_and\_spot})$
    \Else
    \State $\texttt{InstanceManager.free}(-\Delta, \textit{ondemand\_first})$
    \EndIf
    \State $\texttt{ConfigUpdate}(C_{t}, C_{t+1})$
\EndFunction
\end{algorithmic}
\caption{Adaptive configuration optimizer.}
\label{algo:opt}
\end{algorithm}

\Sys uses parallel configurations to identify a strategy to parallelize LLM serving across multiple GPU instances. A {\em parallel configuration} is represented as a tuple $C=(D, P, M, B)$, where $D$, $P$, and $M$ indicate the data, pipeline-model and tensor-model parallelism degrees, and $B$ is the maximum mini-batch size.
A key difference between \Sys and existing spot-instance serving systems is that \Sys can \textit{proactively} adjust its parallel configuration by leveraging the ahead-of-time notifications provided by the cloud to handle instance preemptions and acquisitions.
For each preemption and acquisition notification, \Sys's parallelization controller opportunistically adjusts the parallelization configuration to improve LLM serving performance. Such reparallelization mechanism is also adaptive for fluctuating inference workload, which has been extensively studied in prior approach~\cite{zhang2019mark}.

\paragraph{Grace period of spot instance.}
Modern clouds generally offer a grace period (e.g., 30 seconds on Azure~\cite{azure_spot}) to allow a spot instance to complete running tasks before preempting the instance. 
Allocating new instance doesn't have a grace period, but initializing inference engine also takes a short period of time \revision{(e.g., 2 minutes for launching and initializing in our evaluations)}, which can be measured in advance and treated as the acquisition grace period in \Sys.

\begin{figure*}[t]
    \subfloat[Illustration of configuration update and context migration in \Sys.]{
        \centering
        \includegraphics[width=0.58\linewidth]{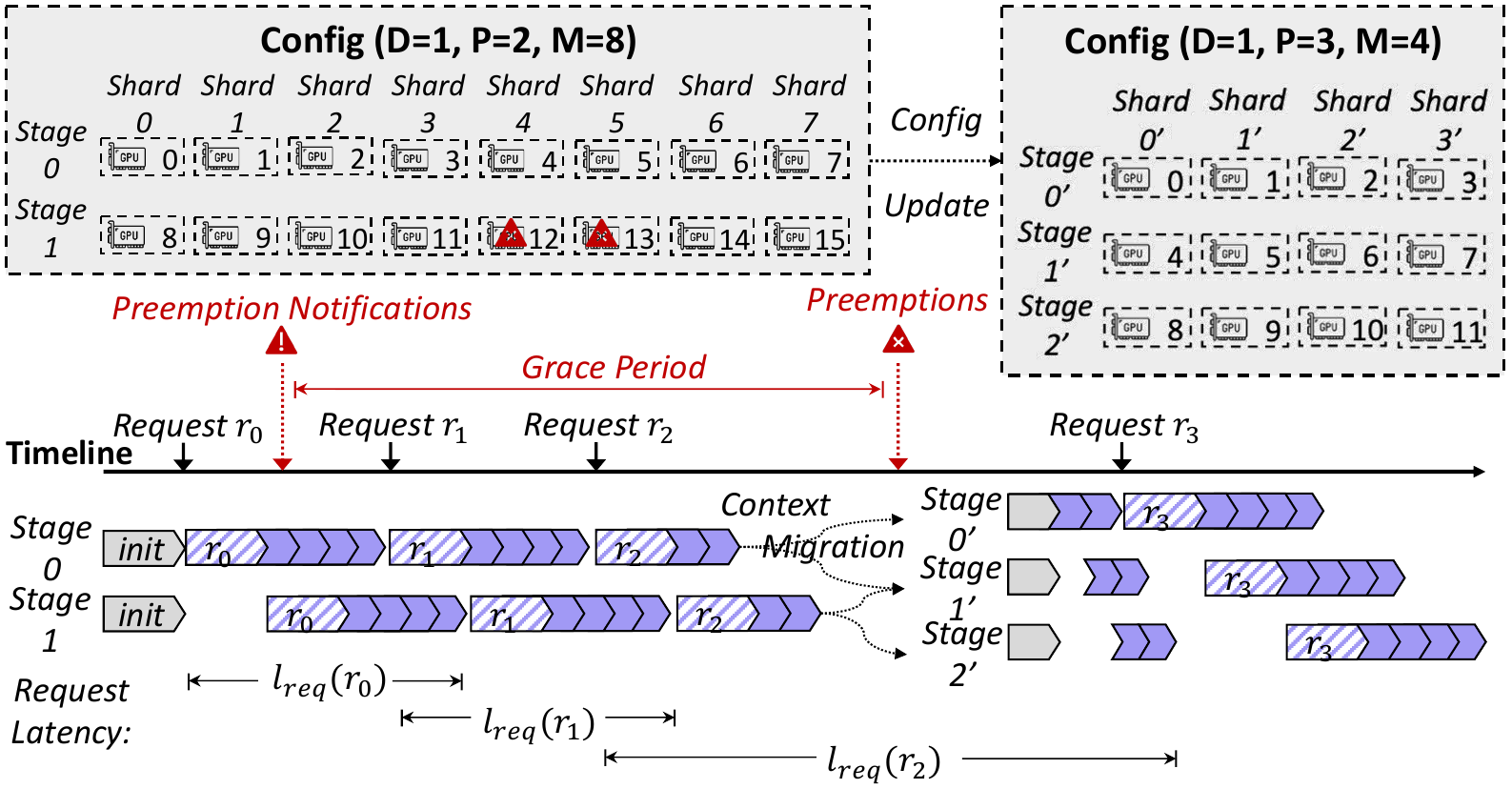}
        \label{fig:reparallel}
    }%
    \subfloat[Illustration of device mapping in \Sys.]{
        \centering
        \includegraphics[width=0.39\linewidth]{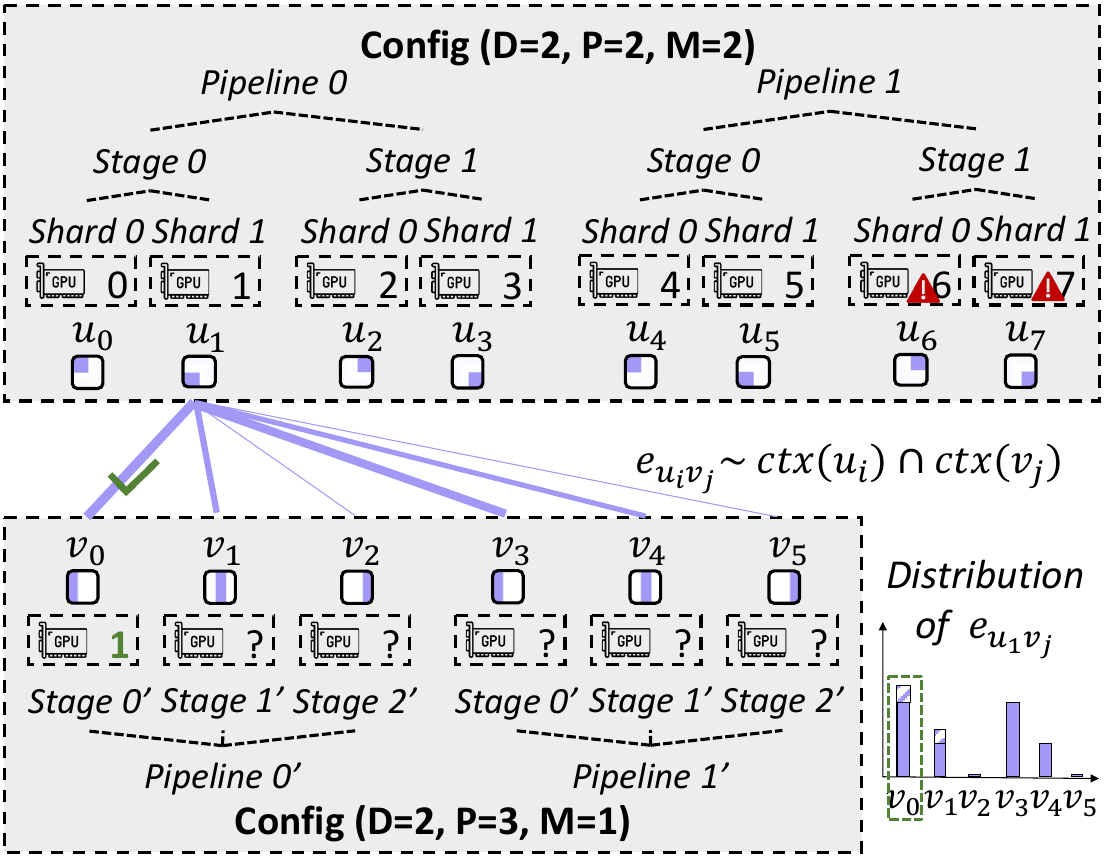}
        \label{fig:mapping}
    }
  \caption{Figure~\ref{fig:reparallel} shows an example of \Sys changes the parallel configuration from (1,2,8) to (1,3,4) through context migration within the grace period and continues previous decoding progress of request $r_3$. Figure~\ref{fig:mapping} shows an example bipartite graph between six available instances (i.e., $u_0\sim u_5$) and topology positions in the new configuration (2,3,1). Here we only draw the weighted edges starting from $u_1$.}
\end{figure*}

\paragraph{Adaptive configuration optimizer.}
\Sys uses an adaptive optimization algorithm to balance the trade-off among throughput, latency, and cost.
We use two time-varying variables $C_t$ and $N_t$ to denote the parallel configuration and the number of available instances at time step $t$. 
Note that $N_t$ considers instances in the grace period, which includes newly allocated instances and excludes instances to be preempted.
Let $\phi(C)$ to be the serving throughput with the parallel configuration $C$ and $\alpha_{t}$  be the request arrival rate at time step $t$\footnote{Since the request arrival rate might change randomly, we estimate $\alpha_{t}$ by observing the request arrivals within a short past duration (e.g., 30s).}.
Algorithm~\ref{algo:opt} shows the workflow of the optimizer, which mainly works when the current serving capability is not compatible with $\alpha_{t}$ due to changes in instances' availiability or serving workload.

Overall, the optimizer minimizes the end-to-end inference latency $l_{req}(C)$ while maintaining a throughput higher than $\alpha_t$ (line 3).
Specially, if there are multiple configurations that can achieve similar minimum inference latency, \Sys selects the configuration with lower monetary cost (i.e., using fewer instances).
Note that, in addition to minimizing $l_{req}(C)$, other targets are also feasible, such as meeting the requirements of pre-defined SLO (i.e., $l_{req}(C)\leq l_{SLO}$).
When \Sys's peak serving throughput can not exceed the request arrival rate $\alpha_t$ (i.e., $\nexists C. \phi(C) \geq \alpha_t$), \Sys updates its parallel configuration to maximize the overall serving throughput (line 5).
The suggested configuration $C_{t+1}$ may  require more or less instances than before (line 6).
Since the allocation of spot instance might not always success, \Sys supports to optionally allocate on-demand instances to further improve serving throughput. 
Specifically, the instance manager allocates on-demand and spot instances at the same time (line 8) to avoid the waiting overhead when spot-instance allocation fails. 
The instance manager is also in charge of releasing the allocated instances (line 10) to alleviate over-provision, where on-demand instances have higher priority due to their costs.
To alleviate the impacts of frequent disturbance of instance availability, \Sys often keeps few addition instances (e.g., two in the experiments of \S\ref{sec:eval}) as a candidate pool for smoother instance substitution.
Finally, \Sys updates the parallel configuration (line 11), and the interruption arranger (\S\ref{sec:inter_infer}) decides when to complete reparallelization, especially for the cases triggered by instance availability changes. This step is still necessary even when $C_{t+1} = C_{t}$, since instance preemptions and acquisitions update instances' memberships.

The optimizer runs online and has negligible overhead (i.e., less than 1s) since the latency estimation of different configurations is done offline in advance. 
\Sys's configuration exploration space is much larger than prior approach like Varuna~\cite{athlur2022varuna} which only considers data and pipeline parallelism. 
It is also possible to extend \Sys to more complicated model parallelisms~\cite{zheng22-alpa,unger2022unity}, which we leave as future work.

\subsection{Device Mapper}
\label{sec:mapper}
Given the target configuration $C_{t+1}$, a straightforward approach to migrating instances is to restart the inference engines on current instances and reinitialize all available GPU instances from scratch. 
%%%
However, this approach does not leverage the opportunity to reuse the model parameters and KV cache available on existing GPU instances, resulting in unnecessary migration cost and inference delay.
As shown in Figure~\ref{fig:reparallel}, instead of destroying and rebuilding these context, \Sys adopts a more lightweight context migration mechanism and can resume interrupted requests' inference.
As migrating these context information among GPU instances may also increase latency, a key challenge \Sys must address is mapping the available GPU instances to the logical device mesh identified by the new parallel configuration to opportunistically reuse previous context.
We ignore batch size and use $(D, P, M)$ to denote a parallel configuration, where $D$, $P$, and $M$ indicate the data, pipeline model, and tensor model parallel degrees.
\Sys binds each GPU instance with a pipeline-stage-shard topology position $(d, p, m)$, which represents the $m$-th shard ($1\leq m\leq M$) of the $p$-th stage ($1\leq p\leq P$) in the $d$-th pipeline ($1\leq d\leq D$).

To switch between different parallel configurations, \Sys formalizes device mapping as a {\em bipartite graph matching} problem, and uses the Kuhn-Munkres (KM) algorithm to find an optimal device mapping that minimizes total data transmission during context migration.

\paragraph{Bipartite graph matching.} \Sys uses a bipartite graph $\m{G} = (\m{V}_a, \m{V}_t, \m{E})$ to describe device mapping, where each node $u \in \m{V}_a$ is a GPU device, each node $v \in \m{V}_t$ represents a pipeline-stage-shard position of the parallel configuration, and a weighted edge $e_{uv}$ ($u\in \m{V}_a, v\in \m{V}_t)$ indicates the amount of {\em reusable} model parameters and key/value cache when mapping GPU $u$ to position $v$ of the parallel configuration.
%%%
As shown in Figure~\ref{fig:mapping}, given the current state of each GPU's context daemon (i.e., organized as $(D=2,P=2,M=2)$) and a target parallel configuration $(D=2,P=3,M=1)$, \Sys builds a complete bipartite graph and computes the edge weight between every $(u,v)$ pair using the size of their intersection contexts.
For example, $u_1$ holds half sharded context of the first stage in the first pipeline, and overlaps the most model context with $v_0$ and $v_3$ since they are in charge of the first stage of the new pipeline. 
Suppose the new pipeline $0^{'}$ inherits the interrupted inference requests from pipeline $0$, we may prefer to match $u_1$ with $v_0$ as it has more cache context to reuse. 
\Sys transforms the optimal device mapping problem to a bipartite graph matching task and uses the KM algorithm to find a maximum-weight match, which maximally reuses the model parameters and KV cache on available GPU instances and minimizes the total data transmission. 

\Sys also considers the cases when each instance has multiple GPUs with higher inter-GPU bandwidth (e.g., NVLink). We facilitate the hierarchical architecture by conducting a two-step matching (i.e., intra-instance and inter-instance) to discover an optimal solution.
\revision{More details can be found in the supplemental material.}

When the new parallel configuration $C_{t+1}$ handles less concurrent inference requests than the original configuration $C_t$ (i.e., $D_t\times B_t\geq D_{t+1}\times B_{t+1}$)\footnote{Recall that in a parallel configuration $C=(D,P,M,B)$, $D$ and $B$ indicate the number of inference pipelines and the batch size of each pipeline, respectively. Therefore, $D \times B$ is the total number of concurrent requests.}, \Sys discards part of the cached results to avoid exceeding the memory capacity of the new parallel configuration.
To minimize the recomputation cost, \Sys keeps the batches of requests with more decoding progresses (i.e., iterations).

\begin{algorithm}[t]
\small
\begin{algorithmic}[1]

\LeftComment{Progressive Migration}
\Function{MigrationPlanner}{context \texttt{ctx}, $\texttt{plan}=[~]$}
    \State \texttt{plan}.append(<migrate, \texttt{ctx.cache}>)
    \State  $\mathbf{O}$ $\gets$ Layer migration order from MemOptMigPlanner
    \For{layer index $i$ in $\texttt{range}$(0, \#layers)}
        \State \texttt{plan}.append(<migrate, \texttt{ctx.weight}$[\mathbf{O}_{i}]$>)
        \State $p \gets$ Get pipeline stage index of layer $\mathbf{O}_{i}$ 
        \If{stage $p$ completes all context migration}
            \State \texttt{plan}.append(<start, instances of stage $p$>)
        \EndIf
    \EndFor
\EndFunction

\LeftComment{Memory Optimized Migration}
\Function{MemOptMigPlanner}{maximum buffer size $U_{max}$}
\State $\mathbf{O} \gets [~]$, $\mathbf{X} \gets \{~\}$
\State Instance buffer memory usage $\mathbf{U}\in\{0\}^N$
\For{layer index $i$ in $\texttt{range}(0, \text{\#layers})$}
\If{(migrate, \texttt{ctx.weight}$[i]$) doesn't exceed $U_{max}$}
\State Update buffer memory usage $\mathbf{U}$
\State $\mathbf{O}$.append($i$)
\Else
\State $\mathbf{X}$.add($i$)
\EndIf
\EndFor
\While{$X$ is not empty}
\State \revision{$x_{opt} \gets$}
\Statex \revision{\hfill $\argmin\limits_{x\in \mathbf{X}} \max\limits_{0\leq i\leq N-1}\{\mathbf{U}_i\mid (\text{migrate}, \texttt{ctx.weight}[x])\}$}
\State $\mathbf{O}$.append($r_{opt}$)
\State $\mathbf{X}$.remove($x_{opt}$)
\EndWhile
\EndFunction
\end{algorithmic}
\caption{Workflow of the \Sys migration planner.}
\label{algo:sched}
\end{algorithm}

\subsection{Migration Planner}
\label{sec:migrate}

After mapping the available devices into the logical parallel positions, the next challenge is to determine the exact migration plan to finish the configuration adjustment.
A naive approach is to make all instances follow a default tensor transmission order and wait until all instances' context are successfully transferred.
This solution mainly has two drawbacks. One problem is that sending all context might be time-consuming especially for large models. To alleviate the context migration overheads, we propose a \textit{progressive migration} schedule that utilizes the pipeline structure and prioritize the migration of front model layers' context. Then the front pipeline stages' instances can start serving, which can be potentially overlapped with the following stages' migration. Ideally, the context migration overheads could be reduced into the cost of a single stage's context transferring. 
Note that, we also prioritize the transfer of all layers' cache context considering the interruption fault-tolerance.
Although it does not achieve the maximum overlapping, it can minimize the possibility of decoding progress lost.

Another problem is the memory consumption of the buffer space for context communication. The migration of every context tensor changes the runtime memory usage. Specifically, the sender's memory can be released while the receivers' memory consumption will increase. An improper migration plan may significantly increase the peak memory usage and leads to sub-optimal inference configurations (e.g., splitting the model into more stages) with higher latency. To provide a memory efficient migration plan, we propose to consider the memory usage during the progressive migration process.
As shown in Algorithm~\ref{algo:sched}, we start from a naive plan in sequential order of the layer index (line 12), applies the context migration of each layer (line 13), and tracks the buffer memory usage of each instance (line 14). The algorithm requires a default hyper-parameter $U_{max}$ to represent the maximum threshold of buffer memory consumption for every instance. We first skip those layers whose context migration might exceed the buffer memory upper bound (line 17). After that, it generates the order of the rest layers by solving a min-max problem (line 19). In particular, it prefers to selects the layer whose context migration can minimize the maximum instance buffer memory usage. In this way, the combined layer context migration order has lower memory consumption and can be used to generate the final migration plan (line 3). 

\section{Stateful Inference Recovery}
\label{sec:inter_infer}
This section introduces {\em stateful inference recovery}, a new inference mechanism that allows \Sys to recover interrupted inference request without recomputation.
In addition, we discuss \Sys's mechanism to handle frequent interruptions.

\subsection{Just-in-time Arrangement}
When instance preemption or acquisition notifications trigger reparallelization, \Sys must decide when to terminate the inference engine and start the context migration for each GPU instance. A conservative approach is to immediately suspend the inference engine to preserve enough time for context migration. However, this approach would interrupt all active requests on the instance.
These unfinished requests must be rerouted and restarted on other inference pipelines, resulting in high end-to-end inference latency.
An aggressive alternative is to finish all active requests first, which might prevent the instance from finishing migration before the preemption.

To avoid these problems, \Sys leverages the grace period offered by the cloud to opportunistically commit inference progress at the token level, which allows an inference request to be interrupted at any incremental decoding iteration.
Since \Sys's context daemon maintains the state (i.e., cache context) of an inference request, the request can be rerouted to another inference pipeline, which can directly continue its inference using the cached state without recomputing previously generated output tokens.

To determine how many iterations to run during a grace period, \Sys adopts \textit{just-in-time (JIT) arrangement} and let the inference engine decide when to stop decoding.
%%%
Specifically, each spot GPU instance includes an \textit{interruption arranger} that receives a notification when a grace period starts. Based on this notification, the interruption arranger checks the remaining time before feeding a new batch of requests into the inference engine.
%%%
Suppose a batch of input requests are ready to serve at time $t$, \Sys determines the number of the decoding iterations $S_t$ differently based on the interruption type. For instance preemption, we have $S_t = \argmax_{0 \leq S \leq S_{out}} \{l_{exe}(S\mid C_t) < T^- - T_{mig} \}$, where $l_{exe}(S\mid C_t)$ is the execution latency for generating $S$ tokens with $C_t$, $T^{-}$ is the remaining grace period for the preemption, and $T_{mig}$ is the cost of migrating instances for reparallelization. For instance acquisition, we also have $S_t = \argmin_{0 \leq S \leq S_{out}} \{l_{exe}(S\mid C_t) \geq T^{+} \}$, where $T^{+}$ is the remaining grace period for the acquisition (i.e., initialization). 
A key difference between these two arrangements is that we maximize the arranged iterations before preemption and minimize that before acquisition. 
The reason is that, unlike in-advance preemption handling, the context migration occurs after instance acquisition. Besides, both cases should also guarantee that the arrangement will not increase the request's latency (i.e., $T_{mig}<l_{exe}(S_t\mid C_t)$).
For example, if the left time is only able to generate few tokens, simply rerouting might be better as it doesn't add context migration overheads, especially when the arrival requests are spare.

\subsection{Interruption Fault-tolerance}
One problem of the recovery approach is that previous arrangements only consider single interruption cases. For multiple consecutive and compact interruptions, their grace periods might overlap with each other and be insufficient to finish the arranged iterations or migrate the context.
Another problem is that if we underestimate the migration costs due to unforeseen reasons (e.g., network vibration), the remaining time might also not be enough for the instance to follow the arrangements.

To build a reliable serving system, \Sys has several fault-tolerance mechanisms to handle the failures.
First, \Sys manages to delay the acquired instance joining and make the arrangements for prior interruptions feasible. Second, if one instance indeed gets preempted before expected, \Sys has to give up the cache context and only migrates the model context with the rest instances. Specially, when all replicas of the same piece of model context are lost due to unexpected failures, the migration can not work and \Sys has to restart by loading weights locally (e.g., disk) or from remote cloud storage (e.g., S3) to fetch the required model parameters.

\section{Implementation}
\label{sec:impl}
We implement the inference server of \Sys in 5.6K LoC in C++ and 2.2K LoC in Python, including three resident processes responsible for request manger, instance manager and meta-context manager respectively.
The generated migration plan is organized in a JSON  format and sent to running instances with a TCP connection.
We build \Sys's inference engine over FasterTransformer~\cite{fastertransformer}, a highly optimized Transformer inference framework built on top of CUDA, cuBLAS~\cite{cublas}, and C++.
We implement our context daemon and interruption arranger inside the inference engine.
Specifically, the memory allocation of model context and cache context in FasterTransformer has been replaced by acquiring the corresponding GPU tensors from the context daemon.
The context migration operations are implemented by the batched asynchronous NCCL send/recv primitives~\cite{nccl}. The context migration requires additional communication buffer space in GPU memory, which is dynamically allocated and released based on the migration plan.
Since the context daemon and FasterTransformer belong to two different process, we involve CUDA Inter-Process Communications (IPC)\cite{cudaipc} to share the context pointers.
To support overlapping in progressive migration, we add a mutex lock to each context tensor to block the inference before its migration is finished. 
We also \revision{design a cost model and} implement a offline profiler over \Sys to estimate the required inference latency, system throughput and the context migration overheads in advance. 
To make the estimation more accurate, we carefully considers the resource under-utilization affects (i.e., GPU, network, PCIe) due to several practical factors (e.g., rarely small batch size, single input token, over-sharded intra-op parallelism, GPU memory accessing, and too small communication data volume) during \revision{cost profiling and modeling}.

%% file: implementation.tex
\section{Evaluation}
\label{sec:eval}
\subsection{Experiment Setup}
\begin{table}[t]
\centering
\caption{Overview of LLMs evaluated.}
\label{eval:models}
\scalebox{0.9}{
\begin{tabular}{ l|llll }
\toprule
Model & Size & min \#GPUs & (P, M) & $l_{exe}(B=1)$ \\
\midrule
OPT-6.7B & 25.0 GB & 4 & (1,4)&  5.447s\\
GPT-20B & 74.5 GB & 12 & (3,4)& 14.373s \\
LLaMA-30B & 111.8 GB & 16 & (2,8)& 17.540s \\
\bottomrule
\end{tabular}
}
\end{table}

\begin{figure}[t]
    \centering
    \includegraphics[width=\linewidth]{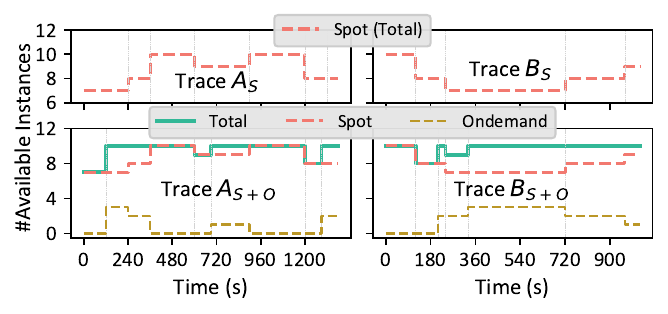}
    \caption{Trace $A_S$ and $B_S$ are extracted from real trace, while $A_{S+O}$ and $B_{S+O}$ are traces created by Algorithm~\ref{algo:opt} mixing on-demand instances based on $A_S$ and $B_S$. Each instance has four GPUs.}
    \label{fig:traces}
\end{figure}

\begin{figure*}[t]
    \centering
    \includegraphics[width=\linewidth]{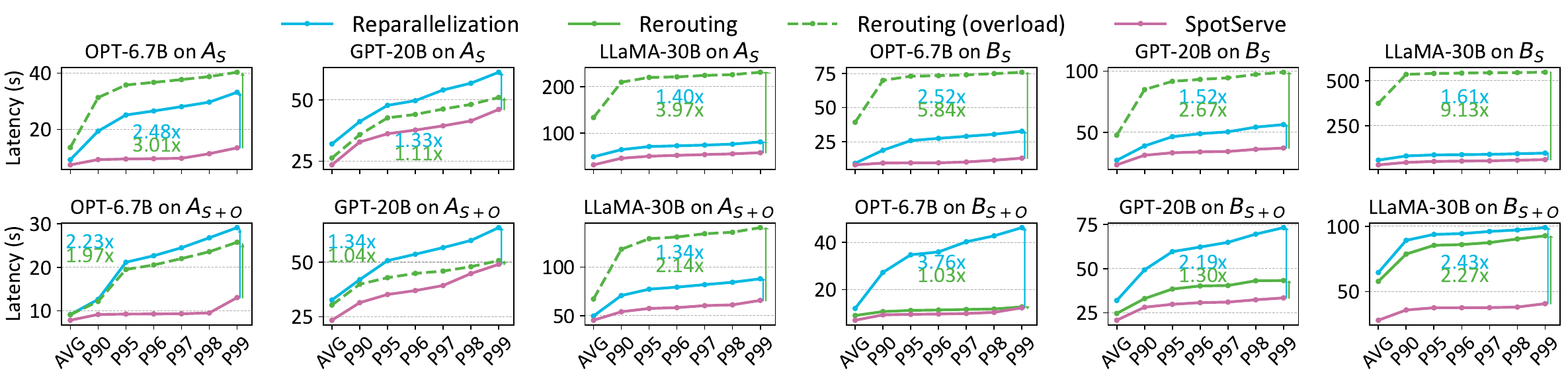}
    \caption{End-to-end serving performance comparison among \Sys, Reparallelization, and Rerouting. The x-axis shows the average and various tail latencies achieved by different approaches, while the numbers report \Sys's latency improvement compared to the baselines.}
    \label{fig:e2e}
\end{figure*}

\begin{figure}[t]
    \centering
    \includegraphics[width=\linewidth]{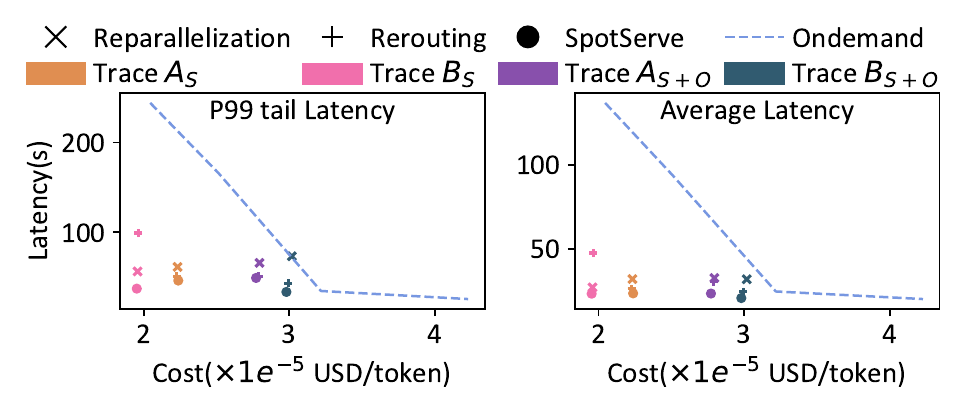}
    \caption{Monetary cost comparison on GPT-20B.}
    \label{fig:price}
\end{figure}

\paragraph{Baseline.}
To our knowledge, \Sys is the first distributed LLM inference system for spot instances.
Therefore, we build two baseline systems on top of FasterTransformer by generalizing two representative ideas of prior approach respectively.
One approach is request \textit{rerouting}, which dynamically reroutes interrupted requests to other available pipelines when preemption happens. It takes a fixed pre-defined optimal model parallel configuration and drops/adds the inference pipeline adaptively. 
Another baseline is model \textit{reparallelization}, which changes the parallel configuration like ours, but has to restart and reinitialize all instance without context migration.
Both of them handle preemption in a reactive manner that has to interrupt the current requests' inference and recompute later. They are implemented with the same inference engine as \Sys to avoid unfairness in the backbone system.
Redundancy-based approaches, which serve several model replicas at the same time, are not included due to the huge cost of LLMs.

\paragraph{Models.} We evaluate \Sys on three LLMs with different scales, including OPT-6.7B~\cite{zhang2022opt}, GPT-20B~\cite{gpt2}, and LLaMA-30B~\cite{touvron2023llama}. Table~\ref{eval:models} summarizes the minimum number of GPUs to serve these models and the corresponding model parallel strategies and their single-request execution latency.

\paragraph{Setting.} We collect a real 12-hour availability trace with AWS \texttt{g4dn} spot instance and extract two representative 20-minute segments (i.e., $A_S$ and $B_S$ in Figure~\ref{fig:traces}) with different dynamic behaviors. For reproducible comparisons, we replay the traces on AWS \texttt{g4dn.12xlarge} instances (4 NVIDIA Tesla T4 GPUs per instance) in our evaluations.
We include both stable and fluctuating inference request arrival workloads.
For static workloads, considering that different models have different computational requirements, we set different request arrival rates for them 
(i.e., 1.5, 0.35 and 0.2 requests/s for OPT-6.7B, GPT-20B and LLaMA-30B by default respectively).
To simulate the bursty in real workloads~\cite{DBLP:journals/corr/abs-2302-11665}, we use Gamma request arrival process with a coefficient of variance (CV) of 6.
Moreover, we separately studied the system performance under the condition of whether to allow mixing with on-demand instances. To achieve that, we generate another two traces (i.e., $A_{S+O}$ and $B_{S+O}$ in Figure~\ref{fig:traces}) following Algorithm~\ref{algo:opt} to opportunistically allocate on-demand instances and mix up with spot instances.
For dynamic workloads, we include a production trace MAF~\cite{MAF-trace} publicly released by Microsoft and discuss in \S\ref{subsec:dynamic-workload}. 
During the optimization, the maximum batch size $B$ is selected from \{1,2,4,8\}, $S_{in}$ is $512$ and $S_{out}$ is $128$.

\begin{figure*}[t]
    \centering
    \includegraphics[width=\linewidth]{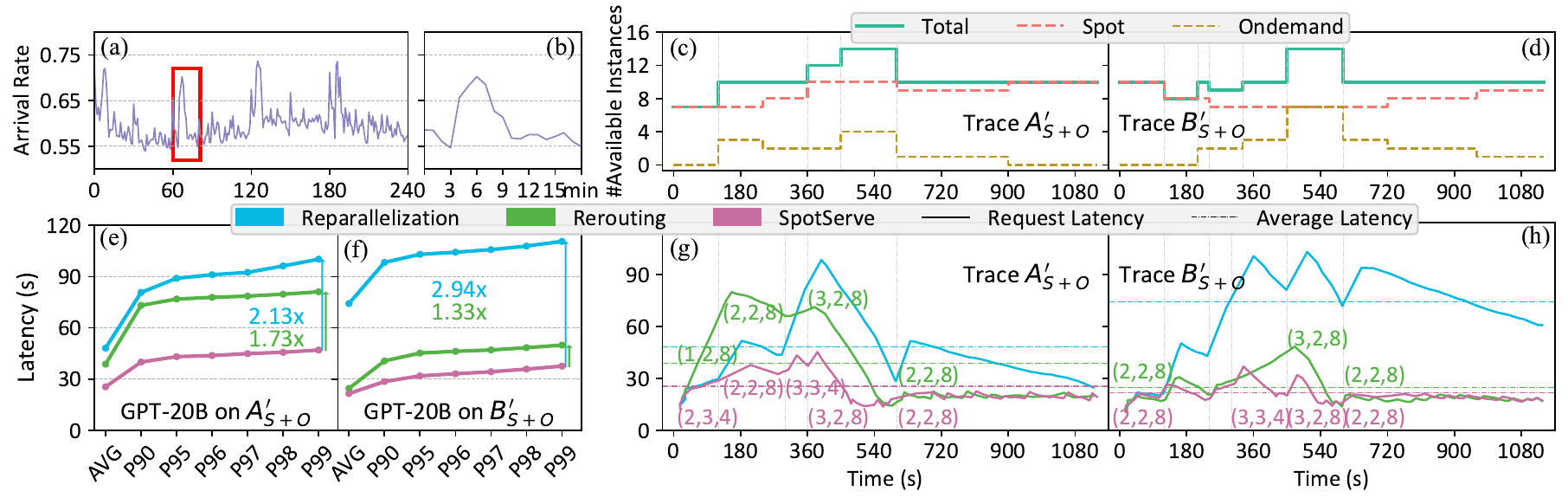}
    \caption{(a) Rescaled MAF trace. (b) The selected trace segment. (c)(d) Two traces based on the  fluctuating workload trace. (e)(f) End-to-end serving performance. (g)(h) Per-request latency throughout the traces, and parallel configurations (D,P,M) after each re-parallelization. Note that the configuration of Rparallelization is always consistent with \Sys. }
    \label{fig:workload-hstack}
\end{figure*}

\subsection{Comparison on Stable Workload}
\label{subsec:e2e_lat}

\paragraph{End-to-end inference latency.}
Figure~\ref{fig:e2e} shows the latency performance of all three models on stable workloads.
\Sys defeats both Rerouting and Reparallelization in terms of all latency metrics on four different traces.  
Taking the P99 latency as an example, \Sys outperforms Reparallelization and Rerouting around 1.34-2.43$\times$ and 2.14-9.13$\times$ respectively on the largest LLaMA-30B model. The improvement mainly comes from three aspects: the dynamic re-paralleization, efficient proactive migration and the stateful inference recovery.

Compared with Reparallelization, the most advantage of \Sys is its lightweight context migration mechanism.
Prior approach like Varuna requires system restarting for each reparallelization and their context has to be lost. Reloading all model parameters and then recompute all interrupted requests will incur long tail latency.

Compared with Rerouting, \Sys can support more fine-grained preemption handling, instead of dropping the entire inference pipeline.
Many cases of Rerouting in Figure~\ref{fig:e2e} are marked with dashed line, representing overload (i.e., the system serving capability becomes lower than the request arrival rate and request accumulation happens).
Taking GPT-20B as an example, when the instance availability is high ($\geq 8$ instances), Rerouting supports a configuration of $(D=2, P=2, M=8)$ with minimum inference latency and sufficient system throughput (i.e., larger than 0.35 requests/s).
Once an instance gets preempted, Rerouting has to drop one inference pipeline and degenerates to $(D=1, P=2, M=8)$, which makes upcoming requests stacked and unable to be served in time.
However, \Sys will serve with $(D=2, P=3, M=4)$ to avoid overload.
% \XM{More severe situation happens in LLaMA-30B where the only inference pipeline is shut down, enforcing Rerouting to stop temporarily and wait for new instances.}
\Sys may occasionally propose the same configuration as Rerouting, but \Sys should still have superior performance because of the KV-cache recovery.
Another observation is that mixing on-demand instances helps alleviate the overload due to the faithful instances acquisitions.

\paragraph{Monetary cost comparison.} Besides inference latency, we also study their monetary cost to see whether it is cost-effective to serving LLM using preemptive instances. Figure~\ref{fig:price} presents the per-token costs of all baseline systems and their latency on GPT-20B model. We also show the results (with the dashed line) of only using on-demand instance, which is more expensive than spot instance (i.e., 3.9 USD/h v.s. 1.9 USD/h).
As the cost decreases, the latency of on-demand instances exhibits a significant increase since it is unable to meet the required serving capability with fewer on-demand instances. In contrast, serving with economical spot instances hit a balance between inference latency and monetary cost.
\Sys significantly saves the cost up to $54\%$ while tolerating a relatively modest increase of less than 18\% in average latency and 90\% in P99 tail latency.
Such cost advantage would be more significant on other types instances with higher on-demand/spot price ratio.

\begin{figure}[t]
    \centering
    \includegraphics[width=\linewidth]{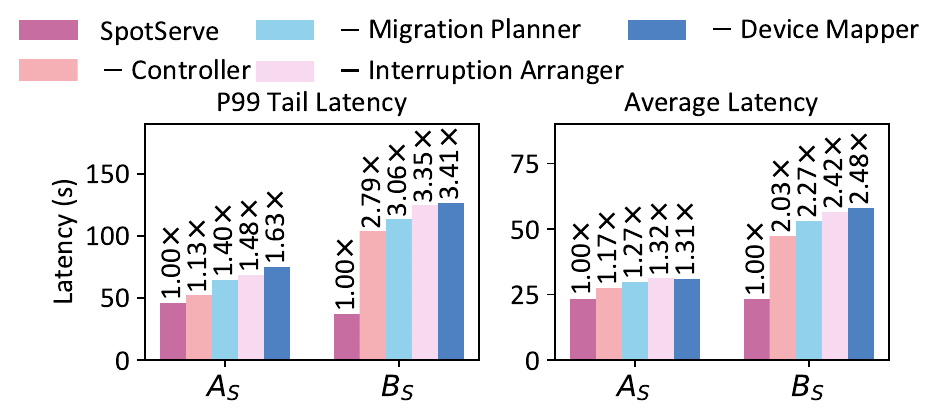}
    \caption{Ablation study of GPT-20B on traces $A_S$ and $B_S$.}
    \label{fig:ablation}
\end{figure}

\paragraph{Ablation study.}
Figure~\ref{fig:ablation} shows the P99 tail latency and average latency of GPT-20B on two traces with different \Sys components. 
We start from \Sys and gradually disables each system optimization one by one. By removing the parallelization controller, the tail latency improves $179\%$ on trace $B_S$. This is because the parallelization controller suggests switching to a new configuration with higher throughput to handle the stacked requests. If we further disables the migration planner, the tail latency improves to $1.4\times$ and $3.1\times$ on traces $A_S$ and $B_S$ respectively. Another important point is that the memory efficient migration planner also reduce the minimum number of GPUs to serve GPT-20B model from 16 to 12, which enlarges the former parallelization configuration exploration space.
The interruption arranger also contributes to $29\%$ tail latency reduction on trace $B_S$ as it transfers the cache context during migration and avoids redundant computation for interrupted requests. Finally, after removing the device mapper, the system degrades to a plain approach only enables model context maintenance without any other optimizations. On the whole, all these optimizations helps \Sys reduces the tail latency by $1.61\times$ on trace $A_S$ and $3.41\times$ on trace $B_S$ respectively.

\subsection{Comparison on Fluctuating Workload}
\label{subsec:dynamic-workload}
To study \Sys's auto-scaling performance, we replay a piece of MAF~\cite{MAF-trace} trace and rescale its arrival intensity like prior approach~\cite{285173,258862,ali2022optimizing} to make it compatible with our experiment setup.
Figures~\sfgref{fig:workload-hstack}{a} and \sfgref{fig:workload-hstack}{b} show that the selected trace includes fluctuating and bursty workload, which is representative in real-world environments.
We enable mixing with on-demand instances in this experiment and the generated instance availability traces (i.e., $A^{'}_{S+O}$ and $B^{'}_{S+O}$) are listed in Figures~\sfgref{fig:workload-hstack}{c} and \sfgref{fig:workload-hstack}{d}.
The end-to-end inference latency statistics are shown in Figures~\sfgref{fig:workload-hstack}{e} and \sfgref{fig:workload-hstack}{f}, and \Sys reduces up to $2.94\times$ and $1.73\times$ P99 tail latency compared with Reparallelization and Rerouting, respectively.

\paragraph{Per-request latency study.} 
Figures~\sfgref{fig:workload-hstack}{g} and \sfgref{fig:workload-hstack}{h} show each arrival request's inference latency over time for both traces.
\Sys almost always performs the lowest latency during the whole trace due to the flexible parallel configuration optimization and the lightweight context migration.
In the following, we take Figure \sfgref{fig:workload-hstack}{h} as an example for in-depth analysis. First, all approaches start with a feasible configuration of $(D=2, P=2, M=8)$ as there are ten spot instances available at $t=0$s.
Preemption first occurs at $t=120$s and $t=240$s but the total available instances are still enough to support $(D=2, P=2, M=8)$.
From $t=270$s, the increasing arrival rate overwhelms the system processing capacity. After 30s, such overload is detected and both \Sys and Reparallelization change the configuration to $(D=3, P=3, M=4)$. The instance acquisition completes at $t=450$s, so they change to $(D=3, P=2, M=8)$ for lower latency. But Reparallelization is suffering from expensive restarting overheads, resulting in the highest peak latency. Rerouting  only changes the number of pipelines and incurs some request waiting overheads.
After $t=600$s, the arrival rate decreasing is detected and on-demand instances start to be released, then both \Sys and Rerouting turn back to $(D=2, P=2, M=8)$. As a result, \Sys significantly outperforms the other two baselines for the fluctuating workloads.

%% file: related_work.tex
\section{Related Work}
\paragraph{DNN inference system.} The widespread DL applications bring great market values and lead to significant DL serving traffics.
Some prior approaches (e.g., Clipper~\cite{201468}, Clockwork~\cite{258862}, Nexus~\cite{shen2019nexus}, and so on) consider temporal multiplexing and increase the GPU utilization through batching and better scheduling. 
INFaaS~\cite{273804} studies the model selection problem when considering multiple models with different inference efficiency or accuracy.
Shepherd~\cite{285173} considers both the resource utilization and the serving system effective throughput and improves the request scheduling.
There are also some inference systems take customized GPU kernel optimization for Transformer models, like TurboTransformer~\cite{DBLP:conf/ppopp/FangYZZ21} and LightSeq~\cite{DBLP:conf/naacl/WangXWWL21}.
Some recent inference systems (e.g., FasterTransformer~\cite{fastertransformer}, Orca~\cite{yu2022orca}, FairSeq~\cite{DBLP:conf/naacl/OttEBFGNGA19}, DeepSpeed~\cite{10046087}, AlpaServe~\cite{DBLP:journals/corr/abs-2302-11665}) support LLM inference by leveraging the parallelization techniques from distributed training approaches. Among them, AlpaServe is designed for resource multiplex scenarios and does not show performance superiority in our empirical study on single LLM inference (i.e., around 3$\times$ lower than FasterTransformer C++ version). Almost all of these prior work are designed for dedicated instances and can not tolerate instance preemptions.

\paragraph{ML Serving over Spot Instance.}
Previous approaches have also involved spot instances into ML inference systems for small ML models. 
Cocktail~\cite{276950} leverages cheap spot instances to increase the number of ensembling models and instance preemptions can lead to certain intermittent loss in accuracy. MArk~\cite{zhang2019mark} studies the over-provisioning problem in previous auto-scaling systems (e.g., SageMaker) for ML serving and improves the cost-effectiveness by using a SLO-aware resource provision algorithm. It also considers involving spot instances for more cost savings \revision{but requires burstable CPU instances to handle the outstanding requests during instance interruptions}. These approaches take a first step to use preemptible instances to serve ML models and motivate our approach on distributed inference of LLMs.

\paragraph{Serverless Computing and ML Serving.}
\revision{There are some recent approaches~\cite{9355312,DBLP:journals/corr/abs-2007-05832,DBLP:conf/usenix/LiZYZL22}} 
applying serverless computing to support ML inference workloads for better cost-effectiveness. However, severless functions are designed to be lightweight with limited computational power, memory and storage, and hard be provisioned with GPUs~\cite{MLSYS2022_0777d5c1}. And serverless functions cannot directly communicate with each other, which is also necessary to support distributed inference of LLMs. As a result, it works well for small models but can not easily serve LLMs due to the hardware constraints.

%% file: conclusion.tex
\section{Limitations and Future Work}
We introduce the limitations of our approach and outline avenues of future research in \Sys. First, the key idea of \Sys is to proactively handle instance availability changes, which strongly relies on the grace period. Although all cloud providers offer this functionality at present, it is still worth exploring more visionary solutions to improve the system performance, such as the combination with inference workload prediction~\cite{285173} or instance availability prediction~\cite{yang2023snape}. Second, our approach mainly focuses on single-type GPU instances. It is also possible to integrate heterogeneous spot instances~\cite{MLSYS2022_0777d5c1} or even instances from different clouds (e.g., SkyPilot~\cite{286502}) for monetary advantages.
These scenarios also bring new challenges to context migration in \Sys. Last, our approach currently takes inference latency minimization as the optimization target. As we mentioned in \S\ref{sec:cfg_opt}, it is still meaningful to explore other targets (e.g., strict SLO~\cite{hao2023deft}, high throughput~\cite{DBLP:journals/corr/abs-2303-06865}) to meet the needs of different inference scenarios. 
Besides, the exploration space of parallelization configurations can be enlarged to support emerging variants of large models (e.g., mixutre-of-experts~\cite{GShard}) in the future.
\revision{While \Sys focuses on spot instances, our techniques can easily generalize to other preemptible resources, e.g., resource scheduler may preempt resources for urgent jobs with switching overheads~\cite{wu2021switchflow}.}
We believe that our approach inspires a new paradigm for distributed inference on preemptible instances, and the insights gleaned from \Sys's design can motivate a variety of following-up research along this direction.

\section{Conclusion}

This paper presents \Sys, the first distributed LLM serving system on preemptible instances.
%%%
Several key techniques in \Sys enable fast and reliable serving of generative LLMs on preemptible instances.
%%%
First, \Sys dynamically adapts the parallelization configuration to make the system serving capability compatible with the workload. The configuration optimization considers the trade-offs among throughput, latency and monetary cost.
Second, to minimize the reparallelization overheads, we design the device mapping algorithm and the migration planning mechanism to achieve efficient context migration.
Finally, to take advantage of the grace period offered by the cloud provider, we introduce stateful inference recovery, which allows \Sys to commit inference progress at a much finer granularity.
We evaluate \Sys on real traces and various scales of popular LLMs and show that \Sys can save 54\% monetary cost compared with on-demand instance and reduce the P99 tail latency by 2.4 - 9.1$\times$ compared with existing approaches.

\section*{Acknowledgement}
We thank the anonymous reviewers and our shepherd, Todd Mytkowicz, for their comments and helpful feedback.
This material is based upon work supported by NSF awards CNS-2147909, CNS-2211882, and CNS-2239351, and research awards from Amazon, Cisco, Google, Meta, Oracle, Qualcomm, and Samsung.

%% file: main_asplos.bbl
\begin{thebibliography}{10}

\bibitem{aws_spot}
Amazon ec2 spot instances.
\newblock \url{https://aws.amazon.com/ec2/spot/}.

\bibitem{nvidia_triton}
Nvidia triton inference server.
\newblock \url{https://developer.nvidia.com/nvidia-triton-inference-server}.

\bibitem{azure_spot}
Use azure spot virtual machines.
\newblock \url{https://learn.microsoft.com/en-us/azure/virtual-machines/spot-vms}.

\bibitem{cudaipc}
Cuda ipc.
\newblock \url{https://docs.nvidia.com/cuda/cuda-runtime-api/group__CUDART__DEVICE.html}, 2021.

\bibitem{fastertransformer}
Nvidia fastertransformer.
\newblock \url{https://github.com/NVIDIA/FasterTransformer}, 2021.

\bibitem{nccl}
Nvidia nccl.
\newblock \url{https://developer.nvidia.com/nccl}, 2021.

\bibitem{vLLM}
vllm: Easy, fast, and cheap llm serving with pagedattention.
\newblock \url{https://vllm.ai}, 2023.

\bibitem{Tensorflow}
Mart\'{\i}n Abadi, Paul Barham, Jianmin Chen, Zhifeng Chen, Andy Davis, Jeffrey Dean, Matthieu Devin, Sanjay Ghemawat, Geoffrey Irving, Michael Isard, Manjunath Kudlur, Josh Levenberg, Rajat Monga, Sherry Moore, Derek~G. Murray, Benoit Steiner, Paul Tucker, Vijay Vasudevan, Pete Warden, Martin Wicke, Yuan Yu, and Xiaoqiang Zheng.
\newblock Tensorflow: A system for large-scale machine learning.
\newblock In {\em Proceedings of the 12th USENIX Conference on Operating Systems Design and Implementation}, OSDI, 2016.

\bibitem{9355312}
Ahsan Ali, Riccardo Pinciroli, Feng Yan, and Evgenia Smirni.
\newblock Batch: Machine learning inference serving on serverless platforms with adaptive batching.
\newblock In {\em SC20: International Conference for High Performance Computing, Networking, Storage and Analysis}, pages 1--15, 2020.

\bibitem{ali2022optimizing}
Ahsan Ali, Riccardo Pinciroli, Feng Yan, and Evgenia Smirni.
\newblock Optimizing inference serving on serverless platforms.
\newblock {\em Proceedings of the VLDB Endowment}, 15(10), 2022.

\bibitem{10046087}
Reza~Yazdani Aminabadi, Samyam Rajbhandari, Ammar~Ahmad Awan, Cheng Li, Du~Li, Elton Zheng, Olatunji Ruwase, Shaden Smith, Minjia Zhang, Jeff Rasley, and Yuxiong He.
\newblock Deepspeed- inference: Enabling efficient inference of transformer models at unprecedented scale.
\newblock In {\em SC22: International Conference for High Performance Computing, Networking, Storage and Analysis}, pages 1--15, 2022.

\bibitem{athlur2022varuna}
Sanjith Athlur, Nitika Saran, Muthian Sivathanu, Ramachandran Ramjee, and Nipun Kwatra.
\newblock Varuna: scalable, low-cost training of massive deep learning models.
\newblock In {\em Proceedings of the Seventeenth European Conference on Computer Systems}, pages 472--487, 2022.

\bibitem{brown2020language}
Tom~B. Brown, Benjamin Mann, Nick Ryder, Melanie Subbiah, Jared Kaplan, Prafulla Dhariwal, Arvind Neelakantan, Pranav Shyam, Girish Sastry, Amanda Askell, Sandhini Agarwal, Ariel Herbert-Voss, Gretchen Krueger, Tom Henighan, Rewon Child, Aditya Ramesh, Daniel~M. Ziegler, Jeffrey Wu, Clemens Winter, Christopher Hesse, Mark Chen, Eric Sigler, Mateusz Litwin, Scott Gray, Benjamin Chess, Jack Clark, Christopher Berner, Sam McCandlish, Alec Radford, Ilya Sutskever, and Dario Amodei.
\newblock Language models are few-shot learners, 2020.

\bibitem{MLSYS2022_0777d5c1}
Junguk Cho, Diman Zad~Tootaghaj, Lianjie Cao, and Puneet Sharma.
\newblock Sla-driven ml inference framework for clouds with heterogeneous accelerators.
\newblock In D.~Marculescu, Y.~Chi, and C.~Wu, editors, {\em Proceedings of Machine Learning and Systems}, volume~4, pages 20--32, 2022.

\bibitem{201468}
Daniel Crankshaw, Xin Wang, Guilio Zhou, Michael~J. Franklin, Joseph~E. Gonzalez, and Ion Stoica.
\newblock Clipper: A {Low-Latency} online prediction serving system.
\newblock In {\em 14th USENIX Symposium on Networked Systems Design and Implementation (NSDI 17)}, pages 613--627, Boston, MA, March 2017. USENIX Association.

\bibitem{cublas}
{Dense Linear Algebra on GPUs}.
\newblock \url{https://developer.nvidia.com/cublas}, 2016.

\bibitem{DBLP:conf/ppopp/FangYZZ21}
Jiarui Fang, Yang Yu, Chengduo Zhao, and Jie Zhou.
\newblock Turbotransformers: an efficient {GPU} serving system for transformer models.
\newblock In Jaejin Lee and Erez Petrank, editors, {\em PPoPP '21: 26th {ACM} {SIGPLAN} Symposium on Principles and Practice of Parallel Programming, Virtual Event, Republic of Korea, February 27- March 3, 2021}, pages 389--402. {ACM}, 2021.

\bibitem{258862}
Arpan Gujarati, Reza Karimi, Safya Alzayat, Wei Hao, Antoine Kaufmann, Ymir Vigfusson, and Jonathan Mace.
\newblock Serving {DNNs} like clockwork: Performance predictability from the bottom up.
\newblock In {\em 14th USENIX Symposium on Operating Systems Design and Implementation (OSDI 20)}, pages 443--462. USENIX Association, November 2020.

\bibitem{276950}
Jashwant~Raj Gunasekaran, Cyan~Subhra Mishra, Prashanth Thinakaran, Bikash Sharma, Mahmut~Taylan Kandemir, and Chita~R. Das.
\newblock Cocktail: A multidimensional optimization for model serving in cloud.
\newblock In {\em 19th USENIX Symposium on Networked Systems Design and Implementation (NSDI 22)}, pages 1041--1057, Renton, WA, April 2022. USENIX Association.

\bibitem{hao2023deft}
Yitian Hao, Wenqing Wu, Ziyi Zhang, Yuyang Huang, Chen Wang, Jun Duan, and Junchen Jiang.
\newblock Deft: Slo-driven preemptive scheduling for containerized dnn serving.
\newblock In {\em Symposium on Networked Systems Design and Implementation}, 2023.

\bibitem{gpipe}
Yanping Huang, Youlong Cheng, Ankur Bapna, Orhan Firat, Dehao Chen, Mia~Xu Chen, HyoukJoong Lee, Jiquan Ngiam, Quoc~V. Le, Yonghui Wu, and Zhifeng Chen.
\newblock Gpipe: Efficient training of giant neural networks using pipeline parallelism.
\newblock In Hanna~M. Wallach, Hugo Larochelle, Alina Beygelzimer, Florence d'Alch{\'{e}}{-}Buc, Emily~B. Fox, and Roman Garnett, editors, {\em Advances in Neural Information Processing Systems 32: Annual Conference on Neural Information Processing Systems 2019, NeurIPS 2019, December 8-14, 2019, Vancouver, BC, Canada}, pages 103--112, 2019.

\bibitem{FlexFlow}
Zhihao Jia, Matei Zaharia, and Alex Aiken.
\newblock Beyond data and model parallelism for deep neural networks.
\newblock In {\em Proceedings of the 2nd Conference on Systems and Machine Learning}, SysML'19, 2019.

\bibitem{kosaian2019parity}
Jack Kosaian, KV~Rashmi, and Shivaram Venkataraman.
\newblock Parity models: erasure-coded resilience for prediction serving systems.
\newblock In {\em Proceedings of the 27th ACM Symposium on Operating Systems Principles}, pages 30--46, 2019.

\bibitem{GShard}
Dmitry Lepikhin, HyoukJoong Lee, Yuanzhong Xu, Dehao Chen, Orhan Firat, Yanping Huang, Maxim Krikun, Noam Shazeer, and Zhifeng Chen.
\newblock Gshard: Scaling giant models with conditional computation and automatic sharding.
\newblock {\em CoRR}, abs/2006.16668, 2020.

\bibitem{DBLP:conf/usenix/LiZYZL22}
Jie Li, Laiping Zhao, Yanan Yang, Kunlin Zhan, and Keqiu Li.
\newblock Tetris: Memory-efficient serverless inference through tensor sharing.
\newblock In Jiri Schindler and Noa Zilberman, editors, {\em 2022 {USENIX} Annual Technical Conference, {USENIX} {ATC} 2022, Carlsbad, CA, USA, July 11-13, 2022}. {USENIX} Association, 2022.

\bibitem{DBLP:journals/corr/abs-2302-11665}
Zhuohan Li, Lianmin Zheng, Yinmin Zhong, Vincent Liu, Ying Sheng, Xin Jin, Yanping Huang, Zhifeng Chen, Hao Zhang, Joseph~E. Gonzalez, and Ion Stoica.
\newblock Alpaserve: Statistical multiplexing with model parallelism for deep learning serving.
\newblock {\em CoRR}, abs/2302.11665, 2023.

\bibitem{liu2021makes}
Jiachang Liu, Dinghan Shen, Yizhe Zhang, Bill Dolan, Lawrence Carin, and Weizhu Chen.
\newblock What makes good in-context examples for gpt-$3 $?
\newblock {\em arXiv preprint arXiv:2101.06804}, 2021.

\bibitem{miao2023galvatron}
Xupeng Miao, Yujie Wang, Youhe Jiang, Chunan Shi, Xiaonan Nie, Hailin Zhang, and Bin Cui.
\newblock Galvatron: Efficient transformer training over multiple gpus using automatic parallelism.
\newblock {\em Proc. {VLDB} Endow.}, 16(3):470--479, 2023.

\bibitem{PipeDream}
Deepak Narayanan, Aaron Harlap, Amar Phanishayee, Vivek Seshadri, Nikhil~R. Devanur, Gregory~R. Ganger, Phillip~B. Gibbons, and Matei Zaharia.
\newblock Pipedream: Generalized pipeline parallelism for dnn training.
\newblock In {\em Proceedings of the 27th ACM Symposium on Operating Systems Principles}, SOSP '19, page 1–15, New York, NY, USA, 2019. Association for Computing Machinery.

\bibitem{pipedream-2bw}
Deepak Narayanan, Amar Phanishayee, Kaiyu Shi, Xie Chen, and Matei Zaharia.
\newblock Memory-efficient pipeline-parallel {DNN} training.
\newblock In Marina Meila and Tong Zhang, editors, {\em Proceedings of the 38th International Conference on Machine Learning, {ICML} 2021, 18-24 July 2021, Virtual Event}, volume 139 of {\em Proceedings of Machine Learning Research}, pages 7937--7947. {PMLR}, 2021.

\bibitem{openai2023gpt4}
OpenAI.
\newblock Gpt-4 technical report, 2023.

\bibitem{DBLP:conf/naacl/OttEBFGNGA19}
Myle Ott, Sergey Edunov, Alexei Baevski, Angela Fan, Sam Gross, Nathan Ng, David Grangier, and Michael Auli.
\newblock fairseq: {A} fast, extensible toolkit for sequence modeling.
\newblock In Waleed Ammar, Annie Louis, and Nasrin Mostafazadeh, editors, {\em Proceedings of the 2019 Conference of the North American Chapter of the Association for Computational Linguistics: Human Language Technologies, {NAACL-HLT} 2019, Minneapolis, MN, USA, June 2-7, 2019, Demonstrations}, pages 48--53. Association for Computational Linguistics, 2019.

\bibitem{gpt2}
Alec Radford, Jeffrey Wu, Rewon Child, David Luan, Dario Amodei, Ilya Sutskever, et~al.
\newblock Language models are unsupervised multitask learners.
\newblock {\em OpenAI blog}, 1(8):9, 2019.

\bibitem{273804}
Francisco Romero, Qian Li, Neeraja~J. Yadwadkar, and Christos Kozyrakis.
\newblock {INFaaS}: Automated model-less inference serving.
\newblock In {\em 2021 USENIX Annual Technical Conference (USENIX ATC 21)}, pages 397--411. USENIX Association, July 2021.

\bibitem{MAF-trace}
Mohammad Shahrad, Rodrigo Fonseca, Inigo Goiri, Gohar Chaudhry, Paul Batum, Jason Cooke, Eduardo Laureano, Colby Tresness, Mark Russinovich, and Ricardo Bianchini.
\newblock Serverless in the wild: Characterizing and optimizing the serverless workload at a large cloud provider.
\newblock In {\em 2020 USENIX Annual Technical Conference (USENIX ATC 20)}, pages 205--218. USENIX Association, July 2020.

\bibitem{shen2019nexus}
Haichen Shen, Lequn Chen, Yuchen Jin, Liangyu Zhao, Bingyu Kong, Matthai Philipose, Arvind Krishnamurthy, and Ravi Sundaram.
\newblock Nexus: A gpu cluster engine for accelerating dnn-based video analysis.
\newblock In {\em SOSP '19: Proceedings of the 27th ACM Symposium on Operating Systems Principles}, pages 322--337, 2019.

\bibitem{DBLP:journals/corr/abs-2303-06865}
Ying Sheng, Lianmin Zheng, Binhang Yuan, Zhuohan Li, Max Ryabinin, Daniel~Y. Fu, Zhiqiang Xie, Beidi Chen, Clark~W. Barrett, Joseph~E. Gonzalez, Percy Liang, Christopher R{\'{e}}, Ion Stoica, and Ce~Zhang.
\newblock High-throughput generative inference of large language models with a single {GPU}.
\newblock {\em CoRR}, abs/2303.06865, 2023.

\bibitem{Megatron}
Mohammad Shoeybi, Mostofa Patwary, Raul Puri, Patrick LeGresley, Jared Casper, and Bryan Catanzaro.
\newblock Megatron-lm: Training multi-billion parameter language models using model parallelism.
\newblock {\em CoRR}, abs/1909.08053, 2019.

\bibitem{DBLP:journals/corr/abs-2007-05832}
Vikram Sreekanti, Harikaran Subbaraj, Chenggang Wu, Joseph~E. Gonzalez, and Joseph~M. Hellerstein.
\newblock Optimizing prediction serving on low-latency serverless dataflow.
\newblock {\em CoRR}, abs/2007.05832, 2020.

\bibitem{piper}
Jakub~M Tarnawski, Deepak Narayanan, and Amar Phanishayee.
\newblock Piper: Multidimensional planner for dnn parallelization.
\newblock In M.~Ranzato, A.~Beygelzimer, Y.~Dauphin, P.S. Liang, and J.~Wortman Vaughan, editors, {\em Advances in Neural Information Processing Systems}, volume~34, pages 24829--24840. Curran Associates, Inc., 2021.

\bibitem{bamboo}
John Thorpe, Pengzhan Zhao, Jonathan Eyolfson, Yifan Qiao, Zhihao Jia, Minjia Zhang, Ravi Netravali, and Guoqing~Harry Xu.
\newblock Bamboo: Making preemptible instances resilient for affordable training of large dnns.
\newblock {\em CoRR}, abs/2204.12013, 2022.

\bibitem{touvron2023llama}
Hugo Touvron, Thibaut Lavril, Gautier Izacard, Xavier Martinet, Marie-Anne Lachaux, Timoth{\'e}e Lacroix, Baptiste Rozi{\`e}re, Naman Goyal, Eric Hambro, Faisal Azhar, et~al.
\newblock Llama: Open and efficient foundation language models.
\newblock {\em arXiv preprint arXiv:2302.13971}, 2023.

\bibitem{unger2022unity}
Colin Unger, Zhihao Jia, Wei Wu, Sina Lin, Mandeep Baines, Carlos Efrain~Quintero Narvaez, Vinay Ramakrishnaiah, Nirmal Prajapati, Patrick~S. McCormick, Jamaludin Mohd{-}Yusof, Xi~Luo, Dheevatsa Mudigere, Jongsoo Park, Misha Smelyanskiy, and Alex Aiken.
\newblock Unity: Accelerating {DNN} training through joint optimization of algebraic transformations and parallelization.
\newblock In {\em 16th {USENIX} Symposium on Operating Systems Design and Implementation, {OSDI} 2022, Carlsbad, CA, USA, July 11-13, 2022}, pages 267--284. {USENIX} Association, 2022.

\bibitem{Transformer}
Ashish Vaswani, Noam Shazeer, Niki Parmar, Jakob Uszkoreit, Llion Jones, Aidan~N. Gomez, Lukasz Kaiser, and Illia Polosukhin.
\newblock Attention is all you need.
\newblock {\em CoRR}, abs/1706.03762, 2017.

\bibitem{DBLP:conf/naacl/WangXWWL21}
Xiaohui Wang, Ying Xiong, Yang Wei, Mingxuan Wang, and Lei Li.
\newblock Lightseq: {A} high performance inference library for transformers.
\newblock In Young{-}bum Kim, Yunyao Li, and Owen Rambow, editors, {\em Proceedings of the 2021 Conference of the North American Chapter of the Association for Computational Linguistics: Human Language Technologies: Industry Papers, {NAACL-HLT} 2021, Online, June 6-11, 2021}, pages 113--120. Association for Computational Linguistics, 2021.

\bibitem{wu2021switchflow}
Xiaofeng Wu, Jia Rao, Wei Chen, Hang Huang, Chris Ding, and Heng Huang.
\newblock Switchflow: preemptive multitasking for deep learning.
\newblock In {\em Proceedings of the 22nd International Middleware Conference}, pages 146--158, 2021.

\bibitem{yang2023snape}
Fangkai Yang, Lu~Wang, Zhenyu Xu, Jue Zhang, Liqun Li, Bo~Qiao, Camille Couturier, Chetan Bansal, Soumya Ram, Si~Qin, et~al.
\newblock Snape: Reliable and low-cost computing with mixture of spot and on-demand vms.
\newblock In {\em Proceedings of the 28th ACM International Conference on Architectural Support for Programming Languages and Operating Systems, Volume 3}, pages 631--643, 2023.

\bibitem{286502}
Zongheng Yang, Zhanghao Wu, Michael Luo, Wei-Lin Chiang, Romil Bhardwaj, Woosuk Kwon, Siyuan Zhuang, Frank~Sifei Luan, Gautam Mittal, Scott Shenker, and Ion Stoica.
\newblock {SkyPilot}: An intercloud broker for sky computing.
\newblock In {\em 20th USENIX Symposium on Networked Systems Design and Implementation (NSDI 23)}, pages 437--455, Boston, MA, April 2023. USENIX Association.

\bibitem{yu2022orca}
Gyeong-In Yu, Joo~Seong Jeong, Geon-Woo Kim, Soojeong Kim, and Byung-Gon Chun.
\newblock Orca: A distributed serving system for {Transformer-Based} generative models.
\newblock In {\em 16th USENIX Symposium on Operating Systems Design and Implementation (OSDI 22)}, pages 521--538, Carlsbad, CA, July 2022. USENIX Association.

\bibitem{zhang2019mark}
Chengliang Zhang, Minchen Yu, Wei Wang, and Feng Yan.
\newblock Mark: Exploiting cloud services for cost-effective, slo-aware machine learning inference serving.
\newblock In {\em USENIX Annual Technical Conference}, pages 1049--1062, 2019.

\bibitem{zhang2019pretraining}
Haoyu Zhang, Jianjun Xu, and Ji~Wang.
\newblock Pretraining-based natural language generation for text summarization.
\newblock {\em arXiv preprint arXiv:1902.09243}, 2019.

\bibitem{285173}
Hong Zhang, Yupeng Tang, Anurag Khandelwal, and Ion Stoica.
\newblock {SHEPHERD}: Serving {DNNs} in the wild.
\newblock In {\em 20th USENIX Symposium on Networked Systems Design and Implementation (NSDI 23)}, pages 787--808, Boston, MA, April 2023. USENIX Association.

\bibitem{zhang2022opt}
Susan Zhang, Stephen Roller, Naman Goyal, Mikel Artetxe, Moya Chen, Shuohui Chen, Christopher Dewan, Mona Diab, Xian Li, Xi~Victoria Lin, et~al.
\newblock Opt: Open pre-trained transformer language models.
\newblock {\em arXiv preprint arXiv:2205.01068}, 2022.

\bibitem{zheng22-alpa}
Lianmin Zheng, Zhuohan Li, Hao Zhang, Yonghao Zhuang, Zhifeng Chen, Yanping Huang, Yida Wang, Yuanzhong Xu, Danyang Zhuo, Eric~P. Xing, Joseph~E. Gonzalez, and Ion Stoica.
\newblock Alpa: Automating inter- and intra-operator parallelism for distributed deep learning.
\newblock In Marcos~K. Aguilera and Hakim Weatherspoon, editors, {\em 16th {USENIX} Symposium on Operating Systems Design and Implementation, {OSDI} 2022, Carlsbad, CA, USA, July 11-13, 2022}, pages 559--578. {USENIX} Association, 2022.

\end{thebibliography}
